\documentclass[debug,overfull]{epl}
\usepackage{amsmath}
\usepackage{amssymb}
\usepackage{amsfonts}
\newcommand{\ud}{\mathrm{d}}

\title{Aubry transition studied by direct evaluation of the modulation 
functions of infinite incommensurate systems}
\shorttitle{Modulation function of infinite systems}
\author{T.S. van Erp \inst{1} \thanks{E-mail:tsvanerp@science.uva.nl} \and A.~Fasolino \inst{2}}
\shortauthor{T.S. van Erp \etal}
\institute{
\inst{1} Department of Chemical Engineering - Universiteit van
Amsterdam, Nieuwe Achtergracht 166, 1018 WV Amsterdam, The
Netherlands\\
\inst{2}
Institute for Theoretical Physics, NSRIM - University of Nijmegen,  
           Toernooiveld 1 - 6525 ED Nijmegen,
           The Netherlands
}

\begin{document}

\maketitle
\begin{abstract}
Incommensurate structures can be described by the Frenkel Kontorova model. 
Aubry has shown that, at a critical value $K_c$ of the coupling 
of the harmonic chain to an incommensurate periodic  potential,
the system displays  the analyticity breaking transition  between 
a sliding and pinned state. The ground state equations coincide 
with the standard map in non-linear dynamics, with smooth or chaotic orbits
below and above $K_c$ respectively. For 
the standard map, Greene and MacKay have calculated the value
$K_c$=.971635. Conversely,
evaluations based on the  analyticity breaking of the modulation function have 
been performed for high commensurate approximants. Here we show how 
the modulation function of the infinite system can be calculated without using 
approximants but by Taylor expansions of increasing order. 
This approach leads to a value $K_c'$=.97978, implying the existence of
a golden invariant circle up to $K_c' > K_c$. 
\end{abstract}

\section{Introduction}
The Aubry transition in the Frenkel Kontorova model is considered 
equivalent to the breakup of tori in the standard map\cite{Floria}. 
Based on Greene's hypothesis that dissolution of invariant tori can
be associated with the sudden change from stability to instability of
nearly closed orbits,
an 
accurate evaluation of the critical coupling $K_c$ has been 
derived \cite{Greene,MacKaythesis}. Besides, by a rigorous computer proof 
\cite{MacKayPer}
an upper limit $K_c < 63/64$ has been established above which no golden
invariant circles can exist.
Here we give a 
comparably accurate estimate of $K_c$ based on the analyticity breaking of
the modulation function. Our approach does not make use of finite commensurate 
approximants, but of Taylor 
expansions up to very high order of the modulation function of infinite systems. 
The fact that the two values of $K_c$ do not coincide 
seems to imply violation of Greene's assumption.

\section{The model}
The Frenkel Kontorova model (FKM) consists of  a harmonic chain of atoms 
interacting with a potential $V$ of period 
incommensurate to the average lattice spacing $l$. 
The total potential energy reads:
\begin{equation}
E=\sum_{i=1}^{N} \frac{1}{2} (x_i-x_{i-1}-l)^2+V(x_i)
\end{equation}
\begin{equation}
\tx{with } V(x)=\frac{K}{(2 \pi)^2}(1-\cos(2 \pi x))
\label{VFK}
\end{equation}
where $x_i$ is the coordinate of particle $i$. The usual choice is 
to set the period of the potential to unity and the average spacing to the
golden mean $l=\tau_g=(\sqrt{5}-1)/2$.

The ground state positions $x_i^G$ can be described 
by the modulation function $g$: $x_i^G=il+g(il)$ where $g(x)$ is periodic 
with the
period 1 of the modulation potential.  

\section{Correspondence with the standard map}

At equilibrium, the force on each particle has to vanish. The relation 
$\frac{\partial E}{\partial x_i}=2x_i-x_{i-1}-x_{i+1}+V'(x_i)=0$ yields 
the standard map.
\begin{equation}
\left( \begin{array}{c} x_{i+1} \\ x_i \end{array} \right)=
\tens{T} \left( \begin{array}{c} x_i \\ x_{i-1}  \end{array} \right)=
\left( \begin{array}{c} 2 x_i + V'(x_i) -x_{i-1} \\
                           x_i          \end{array} \right)
\label{it}
\end{equation}
Starting from a point $(x_1,x_0)$, 
this relation defines iteratively a sequence of points on a torus 
$(x_{i+1}, x_i)$ mod 
$\mathbb{Z}^2$.
For $K<K_c$ subsequent points lie on smooth orbits whereas for $K>K_c$ all 
orbits become chaotic \cite{aubry83,chirikov}. 
The theoretical explanation  of this 
transition follows from the Kolmogorov, Arnold and Moser (KAM) 
theorem\cite{Tabor}.
 
The standard map and the Aubry transition
can be related 
by considering the trajectories with an infinitesimal change in starting point 
and their relation to the sliding phonon mode in FKM.
A small change of starting point $(\delta x_1,\delta x_0)$ will lead to
successive changes from the original trajectory of eq.~(\ref{it}), which
up to first order in $\vect{\delta x}$ are: 

\begin{equation}
\left( \begin{array}{c}
\delta x_{i+1}\\
\delta x_i
\end{array} \right)=
\left( \begin{array}{cc}
2+V''(x_i)&-1\\
1 & 0
\end{array} \right)
\left( \begin{array}{c}
\delta x_{i}\\
\delta x_{i-1}
\end{array} \right)
\label{delx}
\end{equation}

The time dependence of phonon eigenvectors $\vect{\epsilon}$ of the FKM, 
$\epsilon_i  =x_i(t)-x_i^G\sim e^{i \omega t}$, satisfy:
\begin{equation}
\epsilon_{i+1}=2 \epsilon_i-\epsilon_{i-1}+(V''(x_i)-\omega^2)\epsilon_i
\label{eps}
\end{equation}

When $\omega=0$, eq.~(\ref{eps}) for $\epsilon_i$ becomes equivalent to 
eq.~(\ref{delx}) for $\delta x_{i}$. 
If the standard map trajectories are unstable 
with respect to initial conditions, the divergence of $\vect{\delta x}$ 
would imply localization of the sliding mode if $\omega=0$ belongs to the 
spectrum\cite{aubry83}. Since for $\omega=0$,
$\epsilon_i=1+g'(il)$ one can easily imagine that an exponentially localized 
eigenvector yields a discontinuous modulation function $g$. 
This in turn leads to pinning of the FKM and to the disappearance of the
sliding mode. Interestingly we have noted \cite{own1} that in the neighborhood 
of $K_c$ a tendency to localization of 
the lowest frequency mode occurs, showing up in a drop of the participation 
ratio. We have conjectured that localization
of the zero-frequency mode is the precursor of the  structural transition
which leads to the opening of the phonon gap.

The critical value $K_c$ has been determined to a high precision as 
$K_c=0.971635406$ by MacKay by renormalization of invariant
circles within Greene's hypothesis \cite{MacKaythesis}. 
In this paper, we propose a route 
to the calculation of $K_c$ which does not need this hypothesis. 

\section{Modulation functions at the Aubry transition} 
Aubry has discussed the analyticity breaking transition also by 
applying a perturbative approach to the equilibrium equation 
for the modulation function.  
\begin{equation}
2g(x)-g(x+l)+g(x-l)=-V'\big(x+g(x)\big)
\label{mod}
\end{equation}

The KAM theorem could be applied to prove that there exists a continuous 
function,
which obeys eq.~(\ref{mod}), when the potential is smooth enough and $K$ is 
small.   
We follow the empirical approach of ref. \cite{Aubry4}, which gives
insight in the small denominator problem.
By writing $V'$ in Fourier series as
\begin{equation}
V'(x)=K \sum_{m=-\infty}^{+\infty} V_m e^{2 \pi i m x}
\end{equation}
and approximating  $V'(x+g(x)) \approx V'(x)$ in eq.~(\ref{mod}) we find 
the modulation function to be:
\begin{equation}
g(x)=-K \sum_{m} \frac{V_m}{\omega_m^2} e^{2 \pi i m x}
\end{equation}
with $\omega_m=2 | \sin(\pi m l) |$. Notice that this approach applies
to a general periodic potential $V$.
The denominators $\omega_m^2$ can become infinitesimally small, especially for
Fibonacci numbers $F_n (F_0=F_1=1, F_i+F_{i-1}+F_{i-2}$). With the relation
$F_{n-1}-F_n \tau_g=(-\tau_g)^{n+1}$ the divergence of the denominators
are:
\begin{equation}
\lim_{n\rightarrow \infty} \frac{1}{\omega_{F_n}^2}=\frac{1}{4 \pi^2} 
\big(\frac{1}{\tau_g^2}\big)^{n+1}
\label{divom}
\end{equation}

For sufficiently small values of $K$,
this divergence is limited. 
Aubry \cite{Aubry4} gave, with a simple  derivation,
an upper bound
for  $K$ above which an analytical modulation function can no longer exist.
This value for potential (\ref{VFK}) is  $4 \pi$, 
which is much larger than the actual $K_c$.
Our goal is to find the value of $K$ for which divergence 
of the denominators really takes place and compare this value 
with the one obtained by MacKay \cite{MacKaythesis}. 
The equivalence of these two approaches 
relies on the validity of Greene's hypothesis.
\section{Approximations of the modulation function of infinite systems and 
evaluation of $K_c$}

We rewrite eq.~(\ref{mod}) with potential
(\ref{VFK}) by applying
the discrete Fourier transform:
\begin{equation}
g(x)=\sum_{k=-\infty}^{+\infty}
X_k e^{2 \pi i k x} \quad \tx{with inverse:} \quad
X_k=\int_0^1 
\ud x \, g(x) e^{-2 \pi i k x}
\label{Fou}
\end{equation}
yielding:
\begin{equation}
\omega_k^2 X_k= \frac{K i}{4 \pi} \int_0^1 \ud x \, e^{-2 \pi i k x} 
\Big[e^{2 \pi i x} e^{2 \pi i g(x)} - e^{-2 \pi i x} e^{-2 \pi i g(x)} \Big]
\label{expg}
\end{equation}
By expanding $\exp(2\pi i g(x) ) $ as in \cite{Con_prl} eq.~(\ref{expg}) becomes:

\begin{eqnarray}
\omega_k^2 X_k= \frac{K i}{4 \pi} 
\sum_{m=0}^{\infty} \frac{(i 2 \pi)^m}{m!}
\sum_{k_1 k_2 \ldots k_m} \bigg[ X_{k_1}
X_{k_2}\ldots X_{k_m} \delta_{k_1 + k_2 +\ldots + k_m+1, k}
\nonumber \\
- (-1)^m X_{k_1} X_{k_2}\ldots X_{k_m}
\delta_{k_1 + k_2 +\ldots + k_m-1, k} \bigg] 
\label{TayFK}
\end{eqnarray}

To transform eq.~(\ref{TayFK}) into an expansion in $K$ we define: 
\begin{equation}
X_k=K X_k^1+ K^2 X_k^2 + K^3 X_k^3 + \ldots
\label{XTay}
\end{equation}
yielding:

\begin{eqnarray}
\omega_k^2 X_k^{n}= \frac{i}{4 \pi} 
\Big[ \delta_{1,k}-\delta_{-1,k} \Big] \delta_{1,n} +
\frac{i}{4 \pi} \sum_{m=1}^{\infty} \frac{(i 2 \pi)^m}{m!}
\sum_{n_1 n_2 \ldots n_m}  \sum_{k_1 k_2 \ldots k_m} 
\delta_{n_1 + n_2 +\ldots + n_m+1,n}
 \nonumber \\
\bigg[ X_{k_1}^{n_1}
X_{k_2}^{n_2}\ldots X_{k_m}^{n_m} \delta_{k_1 + k_2 +\ldots + k_m+1, k} 
- (-1)^m X_{k_1}^{n_1} X_{k_2}^{n_2}\ldots X_{k_m}^{n_m} 
\delta_{k_1 + k_2 +\ldots + k_m-1, k} \bigg]  
\label{TayFou}
\end{eqnarray}
 
Eq. \ref{TayFou} could have been obtained directly
by applying to  both sides of eq.~(\ref{mod}) 
the operator $D \equiv \frac{1}{n!} \frac{d^n}{dK^n} \int_0^1
\ud x \, e^{-2 \pi i k x} \ldots |_{K=0}$.
From eq.~(\ref{TayFou}) it is straightforward to obtain first order
approximation $g(x)=-\frac{K}{2 \pi} \frac{\sin(2 \pi x)}{\omega_1^2}+\mathcal{O}(K^2)$.
Since higher
harmonics $X_k$ scale with $K^k$ \cite{Con_prl} $X_k^n=0$ for $|k|>n$. 
A manageable iterative algorithm to calculate the 
coefficients $X_k^n$ from those of lower order can be derived by defining the 
matrix $P(n,k,m)$ as: 

\begin{equation}
P(n,k,m)=\frac{(2 \pi i)^m}{m!} \sum_{n_1 n_2 \ldots n_m} \sum_{k_1 k_2 \ldots k_m} 
X_{k_1}^{n_1}X_{k_2}^{n_2}\ldots X_{k_m}^{n_m} \delta_{n_1+n_2+\ldots + n_m,n} 
\delta_{k_1+k_2+\ldots + k_m,k} 
\label{defP}
\end{equation}
$P(n,k,m)=0$
for $|k|>n$ and for $m>n$. Eventually we are only interested in the 
elements with $m=1$
which give the Taylor-Fourier coefficients of the modulation function
\begin{equation}
X_k^n=\frac{P(n,k,1)}{2 \pi i}
\label{XP}
\end{equation}

\begin{figure}
\onefigure[scale=0.7]{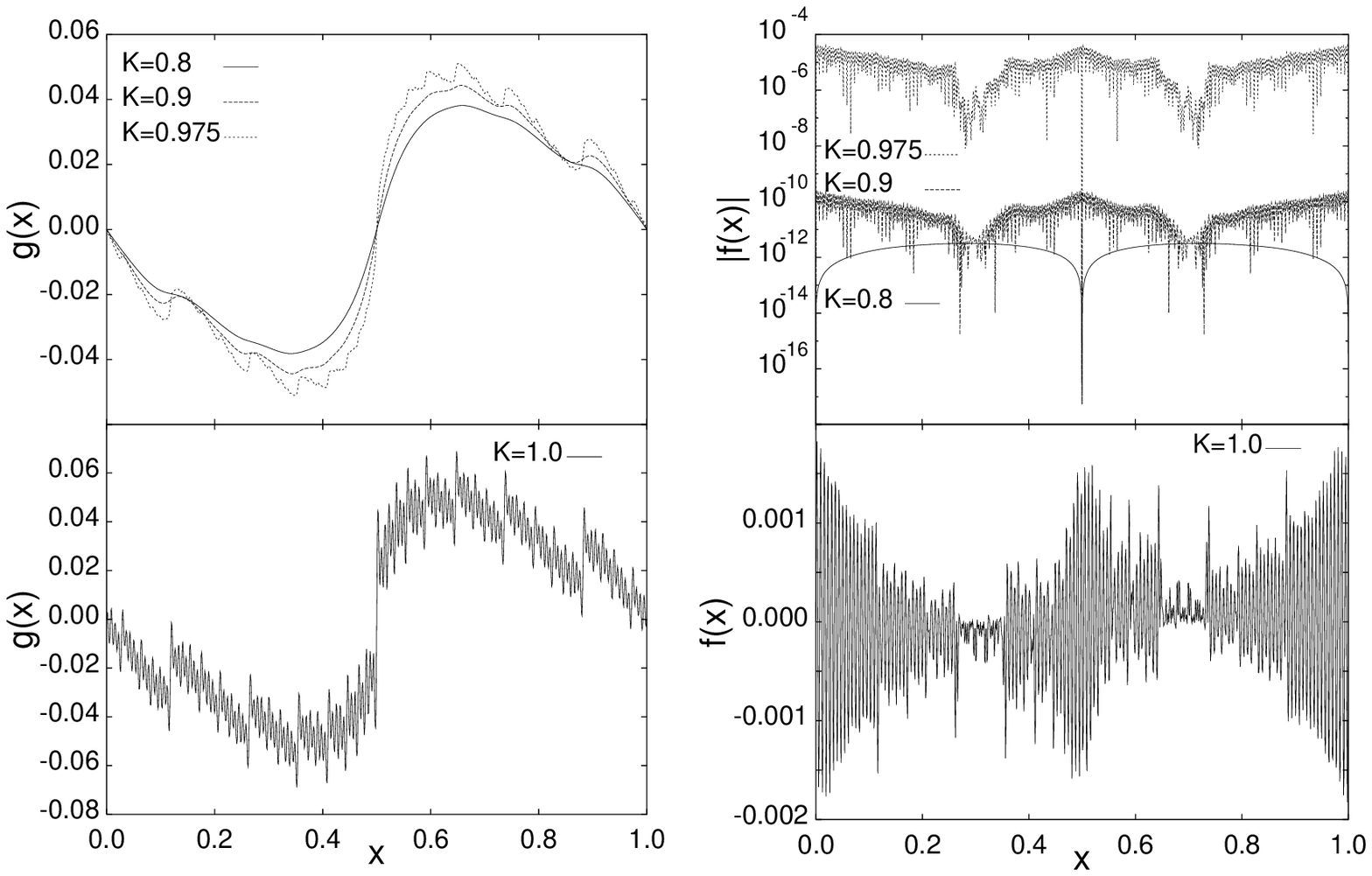}
\caption{Left: $n=150$-th order modulation functions, right: resulting forces\
for the FKM.
Top: $K=0.8,0.9$ and $0.975$, bottom: $K=1.0$}
\label{figmodFK}
\end{figure}

\begin{figure}
\twofigures[scale=.48]{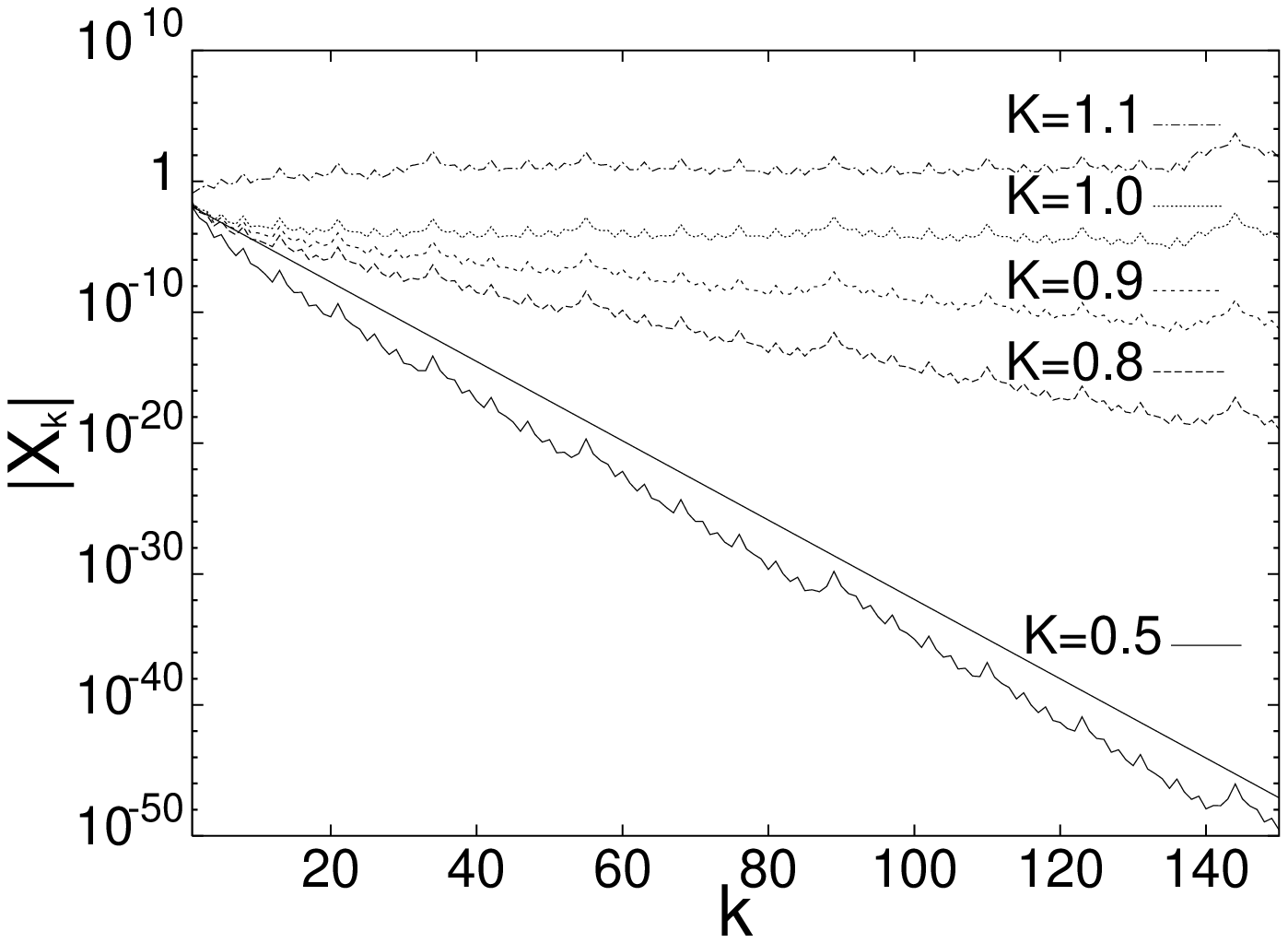}{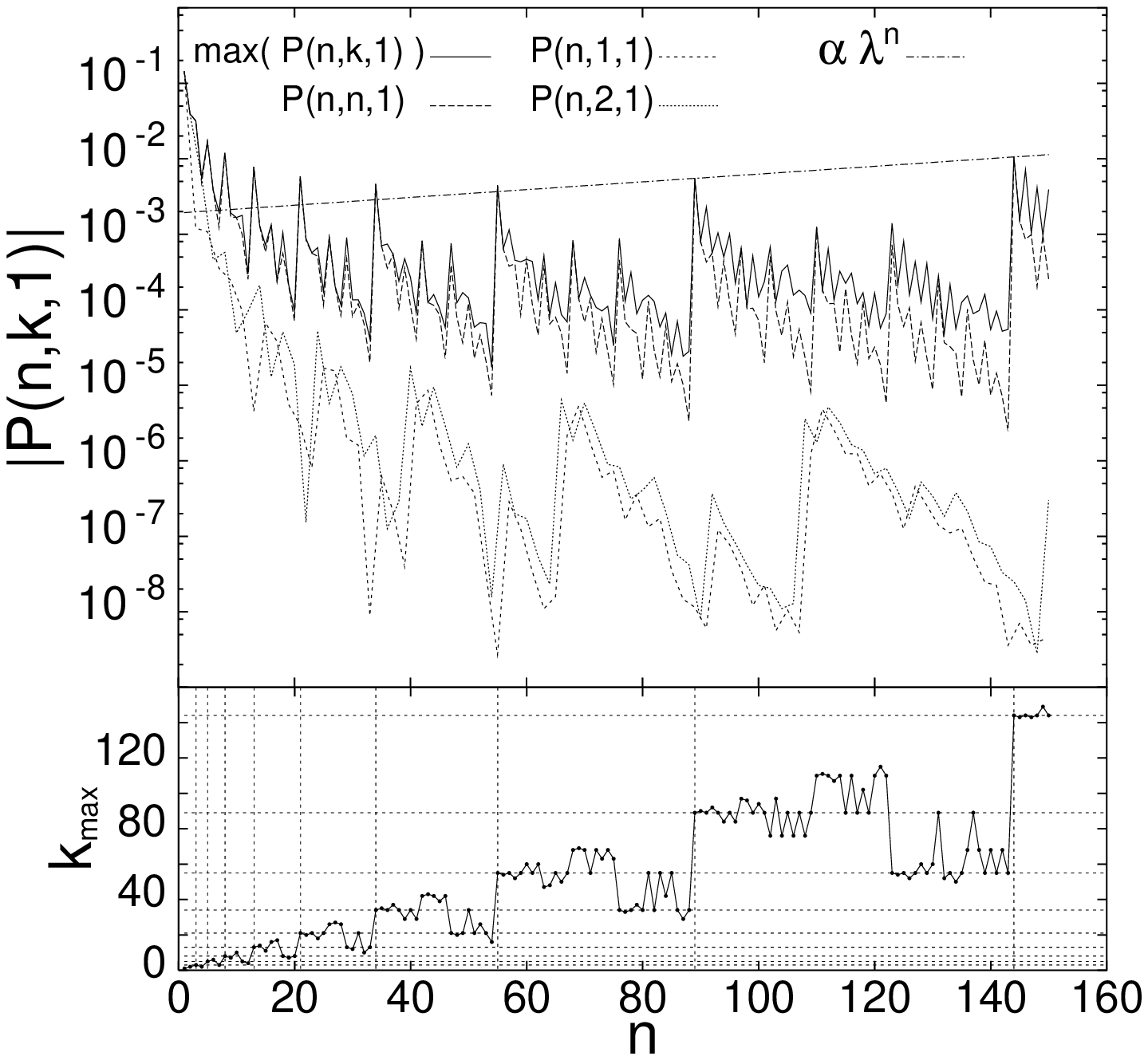}
\caption{$|X_k|$ as function of $k$ for $n=150$ and several values of
$K$ for the
FKM. The expected behavior given by $K^k$ is also given for $K=0.5$.}
\label{figXk}
\caption{$|P(n,k,1)|$ as function of $n$ for different $k$ for the FKM.
The plotted curves are max($|P(n,k,1)|$) (the $k_{max}$ value of $k$
corresponding to this
maximum is given in the lower panel), $|P(n,n,1)|$,$|P(n,1,1)|$ and
$|P(n,2,1)|$. Note that when $n$ is a Fibonacci number,
the maximum is always at $k=n$.  The straight line is the
fit $y=\alpha \lambda^n$ through $P(89,89,1)$ and $P(144,144,1)$,
giving $\lambda=1.012 \Rightarrow
K_c(F_{11}) \sim 0.988$. Dotted lines in the lower panel correspond to the
Fibonacci
numbers.}
\label{PFKM}
\end{figure}

However elements of $P$ with $m>1$ are used during the 
calculation, allowing an effective iterative procedure. 
To this purpose we rewrite  eq.~(\ref{defP}) for $m>1$ as:
\begin{eqnarray}
P(n,k,m)=\frac{2 \pi i}{m} \sum_{n_m} \sum_{k_m} X_{k_m}^{n_m}
\frac{(2 \pi i)^{m-1}}{(m-1)!}
\sum_{n_1 n_2 \ldots n_{m-1}} \sum_{k_1 k_2 \ldots k_{m-1}} \nonumber \\
 \Big[ X_{k_1}^{n_1}X_{k_2}^{n_2}\ldots X_{k_{m-1}}^{n_{m-1}} 
\delta_{n_1+n_2+\ldots + n_{m-1},n-n_m} \delta_{k_1+k_2+\ldots + 
k_{m-1},k-k_m} \Big]
\label{rewrP}
\end{eqnarray}
This leads to our final recursive relations:
\begin{eqnarray}
P(1,\pm 1,1)&=&\frac{\mp 1}{2 \omega_1^2}  \nonumber \\
P(n,k,1)&=& -\frac{1}{2} \omega_k^{-2} \sum_{m=1}^{n-1} \Big[ P(n-1,k-1,m)
-(-1)^m P(n-1,k+1,m) \Big] 
\label{recP}\\
P(n,k,m)&=&\frac{1}{m} \sum_{n'=1}^{n-m+1} \, \sum_{k'=\max\{-n',k-n+n' \} }^
{\min\{n',k+n-n'\}}
P(n',k',1) P(n-n',k-k',m-1) , \quad (n \geq m >1 ) \nonumber
\end{eqnarray}
The second relation is nothing but eq.~(\ref{TayFou}) written in term of $P$ 
for $m=1$; The first one starts the iteration for $n=1$.
The third relation uses eq.~(\ref{rewrP}) to calculate elements with $m>1$. All elements of $P(n,k,m)$ are real and
the set of equations (\ref{recP}) is very efficient
for numerical calculations. Memory  
is the bottleneck at very high Taylor order, 
since the number of 
nonzero elements up to order $n$, which have to be stored, increases 
as $ \sim n^3$. CPU time also increases with $n^3$ but
remains moderate, a 150-th order calculation taking about 15
minutes on a Pentium.\\
The iterative calculation of eq.~(\ref{recP}) has to be performed only once
to get, by use 
of eq.~(\ref{XP}), (\ref{XTay})  and (\ref{Fou}), 
all modulation functions of order 
$n$ for $K \in [0:K_c]$. For $K>K_c$ we expect this method to
break up either because eq.~(\ref{XTay}) 
is no longer 
converging for some
$k$ or because infinite high-frequency oscillations appear in the 
discrete Fourier sum. \\
In fig.~\ref{figmodFK} we plot the modulation functions for different $K$,
together with the forces $f(x)\equiv 2 g(x)-g(x-l)-g(x+l)+V'(x+g(x))$,
which should vanish for the exact modulation function. The appearance of 
high frequency oscillations at $K= 1>K_c$
is clearly visible, yielding non-vanishing forces.
In fig.~\ref{figXk} we show the Fourier coefficients $X_k$
up to Taylor order $n=150$ for several values of $K$ across the transition.
Their values agree with numerical evaluations on commensurate
approximants \cite{Con_prE}.
The behaviour as $K^k$ expected for $K\rightarrow 0$ is still a good 
representation even at $K=0.5\sim K_c/2$. This behaviour is found for all $k$ 
and is not limited to $k\lesssim 20$ by numerical accuracy as found in 
\cite{Con_prE}. 

Inspection of the modulation function as given in fig.~\ref{figmodFK} is not 
enough to pinpoint $K_c$. However the behaviour with increasing $n$ 
of the coefficients $X_k^n$ can lead to a precise evaluation based on 
the following reasoning. By assuming that the Taylor-Fourier coefficients grow
with a power law:
\begin{eqnarray}
X_k^n \sim P(n,k,1) \sim (\lambda_k)^{n} \nonumber \\
X_n^n \sim P(n,n,1) \sim (\lambda_n)^{n}
\end{eqnarray}
both $K^n X_n^n$ and 
the sum (\ref{XTay}) remain convergent up to $K\lambda<1$ whence we can 
estimate $K_c$ as:
\begin{equation}
K_c=\frac{1}{\max\{\lambda_k,\lambda_n\}}
\end{equation}

In fig.~\ref{PFKM} we show the elements $|P(n,k,1)|$ as function of $n$ 
for several values of $k$. We also show max($|P(n,k,1)|$), i.e.
the maximum of $|P(n,k,1)|$
for each $n$. The $k$-value corresponding to this maximum, indicated as 
$k_{max}$, is shown in the
lower panel. In particular for $n$ corresponding to Fibonacci numbers
(34, 55 ,89, 144), $k_{max}$ is equal to $n$. Moreover, at these values of $n$ 
we see sudden jumps due to the fact that $\omega_k^{-2}$ diverges like
eq.~(\ref{divom}). Besides, a power law 
dependence of the maxima at Fibonacci numbers begins to develop for large $n$
as indicated by the straight line.

From these numerical calculations we can conclude that $\lambda_n$ 
is the dominant exponent. Therefore for the calculation of $K_c$ we 
need to calculate eq.~\ref{recP} only for $k=n$ where it takes a simpler form 
\begin{eqnarray}
P(n,n,1) &=& -\frac{1}{2} \omega_k^{-2} \sum_{m=1}^{n-1}  P(n-1,n-1,m)
\\
P(n,n,m)&=&\frac{1}{m} \sum_{n'=1}^{n-m+1}  
P(n',n',1) P(n-n',n-n',m-1) , \quad (n \geq m > 1 )
\end{eqnarray}

\begin{figure}
\oneimage[scale=.38,angle=-90]{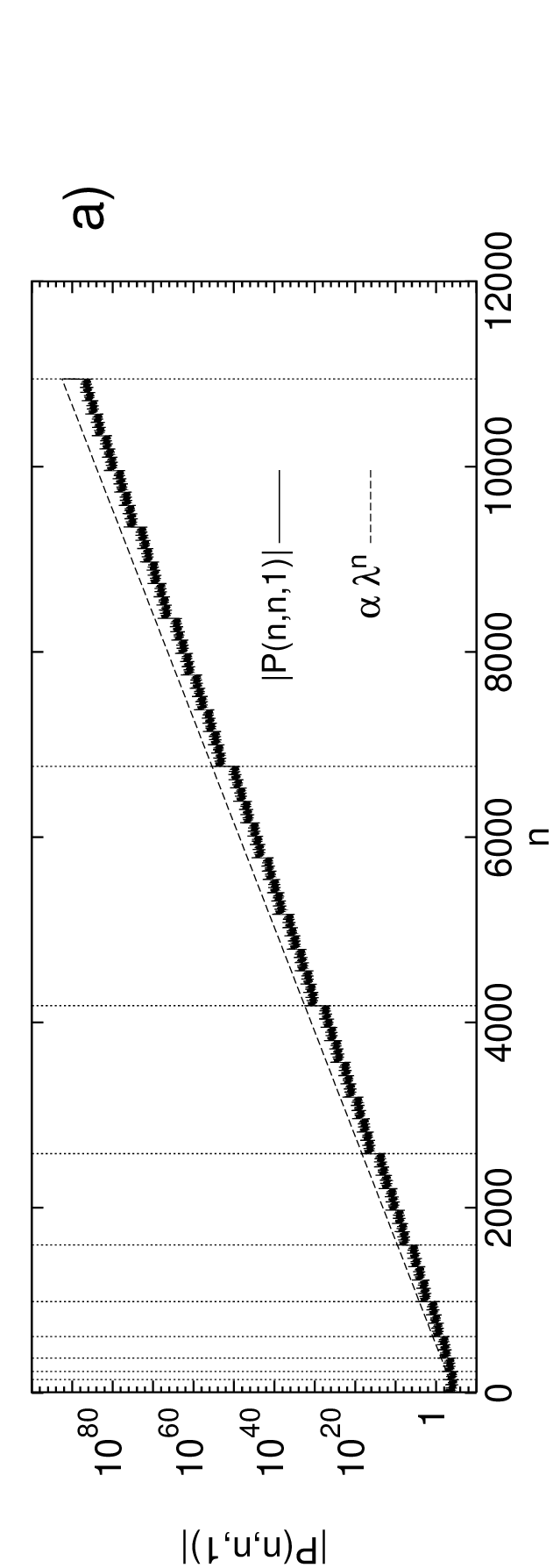}
\oneimage[scale=.38,angle=-90]{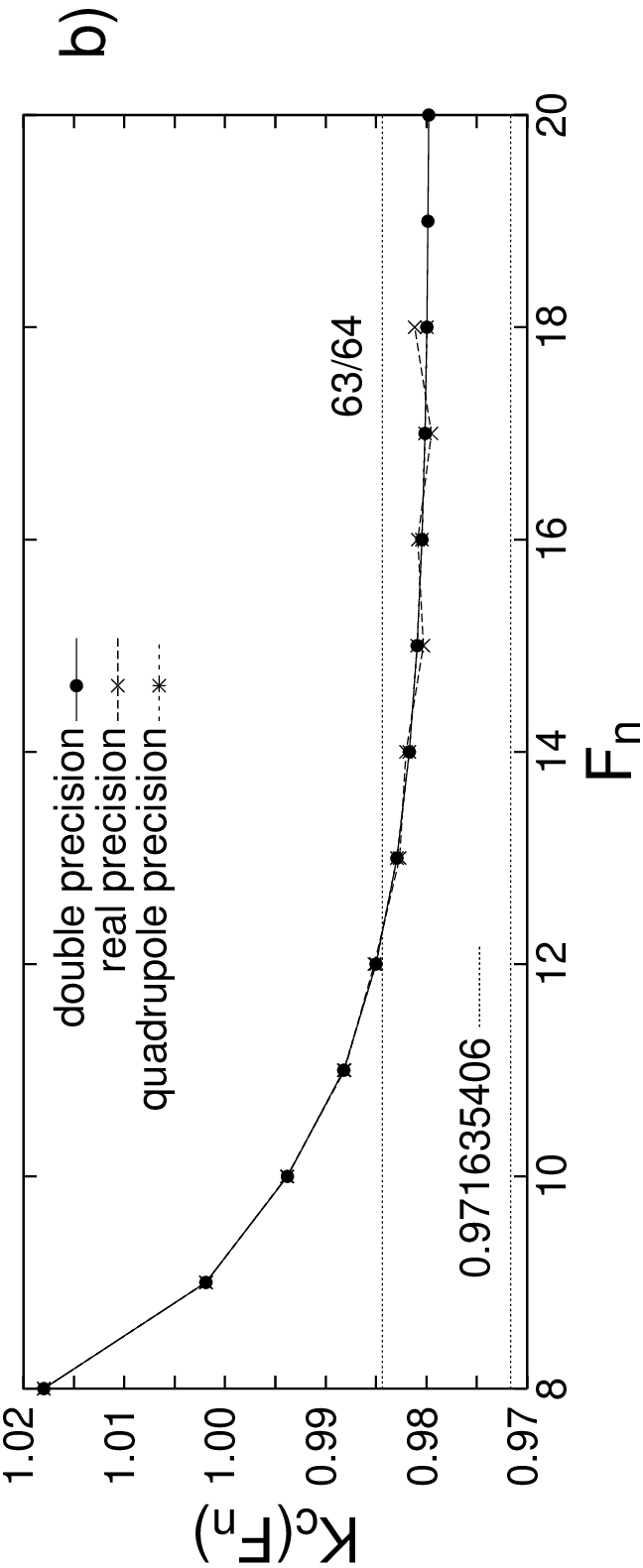}
\caption{a): $|P(n,n,1)|$ as a function of $n$. Vertical lines indicate the
Fibonacci numbers. The calculation is up to order $n=F_{20}=10946$.
The line $\alpha \lambda^n$ corresponds to the best fit through
$n=F_{20}=10946$ and $n=F_{19}=6765$. $1/\lambda$ corresponds to
0.979778542, which is our best approximation of $K_c$.
b):$K_c(F_n)$ approximation from eq.~\ref{eqKcrit} as a function of $F_n$.
Note that the order is up to $F_{20}=10946$. The horizontal dotted
lines are the upper limit $63/64$ \cite{MacKayPer} and MacKay's
value \cite{MacKaythesis}. For examination of floating point errors
also real and quadruple precision
are included up to $F_{18}$. Only real precision gives an
appreciable difference, but
not systematic. }
\label{figPnn1}
\end{figure}

For this case only $\frac{1}{2} n(n+1)$ elements have to be stored,
and calculations up to the order $n$ of a few thousands can easily be achieved
in order to get a good estimate of $\lambda_n$. 
In fig.~\ref{figPnn1}~a) $P(n,n,1)$ is calculated up to $n=F_{20}$ and found 
to have  a power-law behaviour. 
As we assume that $P(F_n,F_n,1) \sim (\frac{1}{K_c})^{F_n}$, we can define
successive approximations of $K_c$  as:
\begin{equation}
K_c(F_n)=\Big| \frac{P(F_n,F_n,1)}{P(F_{n-1},F_{n-1},1)}\Big|^{  \frac{1}{F_{n-2} } }
\label{eqKcrit}
\end{equation}

In fig.~\ref{figPnn1} b) we show $K_c(F_n)$ up to order $F_{20}=10 946$, with
$K_c(F_{20})=0.979778542$, which is close but significantly different than
MacKay's value  $K_c=0.971635406$ \cite{MacKaythesis}.
The value of $K_c=0.979 < 63/64$ determined by our procedure 
seems well converged and is 
unlikely to reach MacKay's value for higher $F_n$. We have also checked that 
quadruple precision does not change our numerical results.   
The discrepancy between our value of $K_c$, which is directly related to 
non analyticity of the modulation function, and that of the 
transition to instability in 
the standard map
questions the validity of Greene's hypothesis.

\end{document}